\documentclass{aastex631}

\def\bfc{}

\def\cm3{cm$^{-3}$}

\def\apjl{ApJL}

\def\aap{A\&A}

\usepackage{graphicx}
\usepackage{epsf}
\usepackage{natbib}

\newcommand{\simlt}{\lower.5ex\hbox{$\; \buildrel < \over \sim \;$}}

\begin{document}

\title{The Distribution of Molecular Water in the Lunar South Polar Region
 based upon 6-micron Spectroscopic Imaging 
}

\shorttitle{Wet Regions on the Lunar Surface }

\author[0000-0001-8362-4094]{William T. Reach}
\affil{Universities Space Research Association, NASA Ames Research Center, Moffett Field, CA 94035, USA}
\email{wreach@sofia.usra.edu}

\author{Paul G. Lucey}
\affil{Hawai'i Institute of Geophysics and Planetology, University of Hawai'i at Mānoa, Honolulu, HI, USA}

\author{Casey I. Honniball}
\affil{NASA Goddard Space Flight Center, Greenbelt, MD, USA}

\author{Anicia Arredondo}
\affil{Southwest Research Institute, San Antonio, TX, USA}

\author{Erick R. Malaret}
\affil{Applied Coherent Technology, Herndon, VA USA}

\begin{abstract}
The amount and distribution of water on the lunar surface are related to the input and production of water by solar wind and meteoroid bombardment, balanced by photodestruction and mobility across the surface. 
Using the Stratospheric Observatory for Infrared Astronomy (SOFIA), we imaged the 6.1 $\mu$m feature that uniquely traces molecular water, covering 1/4 of the lunar nearside surface south of -60$^\circ$ latitude with 5 km resolution on 2022 Feb 17 UTC. 
The water feature strength varies significantly across the region, being drier at +28$^\circ$ longitude to more wet (~170 ppm) at -7$^\circ$ longitude, and also decreasing toward the pole. Significant local enhancements are found, associated with south-facing, high-altitude topographic features. This includes relatively high H$_2$O concentration in a `wet ridge' just north of Curtius {\bfc crater}; the south-facing, northern, inner rims of most prominent craters; the south face of the central peak of Moretus crater; and permanently-shadowed polar regions.
\end{abstract}

\section{Introduction}

The question of whether water is present on the Moon and how it might arise has long been a subject of debate \citep{arnold_ice_1979,crotts_water_2011}. Most lunar rocks and minerals contain very little hydrogen \citep{albarede_intrinsic_2015}, though there are notable exceptions that are of high interest \citep{mccubbin_magmatic_2015}. The lunar surface is covered by regolith, a fragmental layer of debris formed by repeated meteorite impact \citep{mckay_lunar_1991}, which is exposed to hydrogen from the space environment. The solar wind implants protons into the lunar surface, some of which can be retained as molecular hydrogen, hydroxyl, or molecular water \citep{zeller_proton-induced_1966,housley_origin_1973}. The Moon is also bombarded by water-bearing meteorites, asteroids and comets, all of which contribute a portion of their water to the lunar environment \citep{arnold_ice_1979,watson_behavior_1961}. 
The hydrogen content of the lunar regolith outside the polar regions measured by orbital neutron spectroscopy is on the order of 100 $\mu$g/g \citep{lawrence_global_2022}, consistent with laboratory measurement of samples returned by {\bfc the} Apollo {\bfc landing missions}
\citep{warren_optical_2008}. Space-based and {\bfc ground-based} measurements show the lunar surface exhibits absorption bands in the UV and at 3 $\mu$m from high latitude, low-temperature regions 
(\citealt{sunshine_temporal_2009,pieters_character_2009,hendrix_diurnally_2019,chauhan_unambiguous_2021,honniball_telescopic_2020};
see recent review by \citealt{lucey_volatile_2022}).
At high latitudes the abundances of water or hydroxyl inferred from {\bfc near-}infrared measurements are consistent with the abundance of hydrogen estimated from orbital neutron measurements. 
But because both hydroxyl and water have strong absorption at these wavelengths, their relative abundances cannot be separated. 

Water ice has been detected in permanently shadowed portions of the lunar polar regions that stay colder than 100 K \citep{paige_diviner_2010}. The {\bfc Lunar Crater Observation and Sensing Satellite} (LCROSS) mission impacted an expended rocket booster to excavate buried ice at the shadowed polar crater Cabeus, where they found evidence of ice in the ejecta 
{\bfc at an abundance of several percent by weight} 
\citep{colaprete_detection_2010}. Indications from the neutron experiment on Lunar Prospector showed hydrogen is concentrated in the polar regions \citep{lawrence_improved_2006}. Near-infrared reflectance measurements from the Moon Mineralogy Mapper (M3) instrument on the Chandrayaan-1 spacecraft and UV observations by the Lunar Reconnaissance Orbiter showed spectral features consistent with water ice  in permanently shadowed portions of craters near the lunar poles \citep{li_direct_2018,hayne_evidence_2015}. 

Water ice rapidly sublimates at the temperatures of solid bodies 1 AU from the Sun \citep{andreas_new_2007,huebner_photoionization_2015}, so it will not persist from the primordial material from which the Moon originated nor from an Earth-Moon separation event. 
The lunar volatiles, including hydroxyl, ice, and molecular water, are likely delivered from the solar wind, comets, or volcanism \citep{arnold_ice_1979}. By extension from the Moon, other airless bodies may contain exogenic water. The presence of water on Mercury \citep{butler_mercury_1993,lawrence_evidence_2013} and Ceres \citep{kuppers_localized_2014} shows how condensed volatiles can even exist relatively close to the Sun. Constraining the lunar water origin and survival will allow informed hypotheses about whether asteroids, dwarf planets, or exoplanets may have surficial H$_2$O in permanently shadowed regions generated by surface depressions, resonantly locked orbits, and/or tilted rotation axes.

Water molecules were recently detected even on the sunlit surface of the Moon \citep{honniball_molecular_2021,honniball_regional_2022}, and this paper extends those results to present the first well-sampled, wide-area images of localized variations in water abundance. A 6.1 $\mu$m spectral feature uniquely detects H$_2$O (as opposed to just the O-H bond) because it arises from an H-O-H bending mode that requires all 3 atoms of the molecule. 
This feature was detected at a sunlit region near Clavius crater \citep{honniball_molecular_2021} and has been confirmed over a larger area with more recent measurements \citep{honniball_regional_2022}. In this work, we present the results for part of the southern polar region, in order to determine whether the 6.1 $\mu$m water feature correlates with other localized structures. The region near the lunar south pole is of high importance for future space exploration, with multiple spacecraft preparing to go there, including the soon-to-launch NASA Volatiles Investigating Polar Exploration Rover (VIPER) mission \citep{colaprete_volatiles_2020}.

\section{Observations}

\subsection{Observing Method}
Observations were made on 2022 Feb 17 (7:57 to 10:07 UTC) using the Stratospheric Observatory for Infrared Astronomy \citep{young_early_2012}, which flew a 2.5-m effective aperture telescope just above the tropopause and 99.9\% of terrestrial water vapor. The science instrument was the Faint Object infraRed CAmera for the SOFIA Telescope {\bfc (FORCAST)}, with its 256$\times$256 pixel Si:As short-wavelength camera \citep{herter_first_2012}. A 2.4$''$ wide slit and grism dispersing element illuminated the detector with emission from the sky covering 5.1—8.0 $\mu$m in wavelength and 191$''$ in space. 
The secondary mirror oscillated between the target position and a reference position with an amplitude of 180$''$ every 0.4 seconds with a nearly square-wave pattern. The detector integrated signal when the secondary mirror directed the line of sight to the target, then again (with signal multiplied by -1) when the secondary mirror directed the line of sight to the sky reference. For each pointing, 8 seconds of on-source integration (and another 8 seconds off source) was used. The telescope was then moved 600$''$ away, off the southern edge of the Moon, and an identical observation was made. The basic observing sequence comprised tiles with 8 such chop-nod observations, with the slit offset by its 2.4$''$ width, to build a fully-covered map tile of 19$\times$190$''$. 

{\bfc To calibrate systematic effects in the FORCAST spectra, 
water estimates are relative to standard reference sites.  Results from
M3 and ground-based observations \citep{li_direct_2018,honniball_telescopic_2020}
suggest that low-latitude, high-temperature locations should host the
least water, based on the intensity of the 3 $\mu$m water/hydroxyl feature.
For this work we used reference sites in equatorial Mare Fecunditatis
and Oceanus Procellarum.
After each map tile, the reference field  in Mare Fecunditatis  was observed for 5 exposures. 
In the second half of the observations, {\bfc an additional} field {\bfc with low H$_2$ abundance} in Oceanus Procellarum was also observed, as an extra reference. }

A nearby star was also observed after each map tile, as a pointing reference, in order to remove drift of the telescope relative to its nominal pointing; this drift accrues over time when unguided observations, such as our lunar maps, are made. Nine adjacent sequences of tile+reference+star observations were performed, starting from the East, to fully cover the observed region. The first 6 tiles were observed from altitude 40,000 feet (12.2 km), then the aircraft climbed during the 7$^{\rm th}$ tile, with tiles 8 and 9 at 43,000 feet (13.1 km).

\subsection{Data Processing}
All of the raw data were processed using the SOFIA/FORCAST pipeline in the software package “redux” (version 2.10.3), to obtain calibrated, spatially- and spectrally-rectified two-dimensional images for each chop-nod observation {\bfc as is done for all archived SOFIA science observations} \citep{herter_first_2012,clarke_redux_2015}. 
The telluric features due to small amounts of water vapor above SOFIA were removed using an atmospheric transmission model generated using ATRAN \citep{lord_new_1992}. {\bfc The data processing that improves the telluric correction and calibration for these
observations is described in detail by} \citet{arredondo_sofiaforcast_2023} and is summarized here.
A suite of ATRAN models spanning the range of likely precipitable telluric water vapor was generated, then the model that best minimizes the telluric features in the spectra was applied. 
To further remove residual atmosphere and instrumental response, we generated a correction image using reference observations:  
we averaged them together, then for each row (spectrum), a smooth function (2$^{\rm nd}$ order polynomial) was fitted and divided out. The resulting reference calibration image is unit-normalized and includes residual atmospheric emission, evidenced by peaks at the location predicted by ATRAN (but with relative strengths different from the model), as well as uncalibrated instrumental pixel-to-pixel flat-field response. The reference calibration spectral image also contains the average emissivity spectrum of Mare Fecunditatis. 

We divided each on-target chop-nod observation by the reference calibration spectral image. The result comprises the surface brightness spectrum on that part of the Moon divided by the average emissivity spectrum of the reference. 
Thus, all of our measurements are {\bfc relative to the reference}, and the water features are the ratio of water features in the target site to any water features that may be present in the reference site. This means that if there is water at Mare Fecunditatis, we will underestimate the absolute amount of water at the target position. The lunar time of day at the reference was 16:15, and evidence from 3 $\mu$m OH+H$_2$O observations suggests that while the lunar maria are themselves are drier than other places, there does tend to be hydration (though not necessarily molecular water) at that time of day. When observing other relatively dry areas, a zero or even negative spectral feature can be anticipated due to our calibration procedure, for regions that are as dry or drier than the reference. We analyzed the SOFIA Mare Fecunditatis observations using Oceanus Procellarum, where the lunar time of day was 9:05 and very likely dry \citep{honniball_telescopic_2020}, as reference. There is no strong signal from water in Mare Fecunditatis relative to Oceanus Procellarum.

\begin{figure}
\plotone{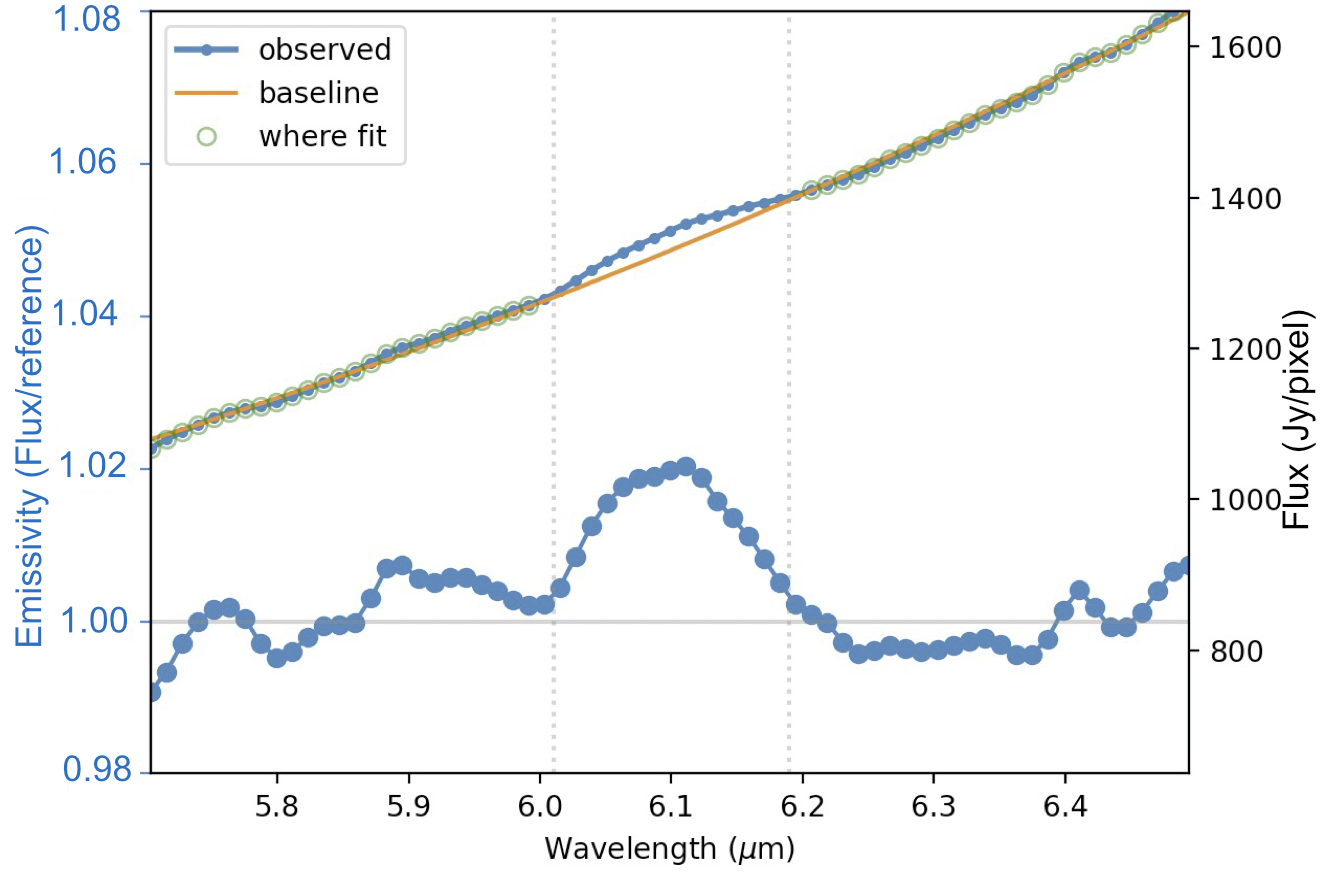}
\caption{Spectrum of the {\bfc 6.1} $\mu$m line from a single pixel {\bfc of the spectral cube}. The upper curve with small blue dots is the reference-corrected spectrum. The orange curve is the baseline fit, with green circles indicating where the baseline was fitted. These curves use the right-hand flux scale. The lower curve with large blue dots is the baseline-divided spectrum, and vertical dotted gray lines show the wavelengths that were summed to generate the H$_2$O image. }\label{fig:specsample}
\end{figure}

The calibrated and reference-corrected images were used to extract the 6.1 $\mu$m emissivity feature of H$_2$O. A local 2$^{\rm nd}$ order polynomial baseline was fitted to each of these spectra, from 5.7 to 6.5 $\mu$m (excluding $\pm$0.1 $\mu$m centered on 6.1 $\mu$m); the spectra were then divided by this baseline. 
{\bfc A first-order baseline generates an almost indistinguishable map of the integrated water strength, but the spectra with only a first-order baseline suffer from artifacts due to the inability to fit the curvature of the thermal continuum (Wien portion of the Planck function) over the fitting window, artificially suppressing the amplitude of the feature.}
Figure~\ref{fig:specsample} shows a spectrum and baseline fit for an individual pixel 
{\bfc of the spectral cube} with a bright H$_2$O feature. 
All spectra were combined to generate a spatial-spectral data cube, using the “redux” software \citep{clarke_redux_2015} and taking into account the relative lunar motion between ephemerides used at the telescope for each tile. We made cubes for the calibrated spectra, then {\bfc for} the reference-corrected spectra,  {\bfc and finally for} the residual (baseline-removed) spectra. The cubes have 276$\times$396 pixels of 0.768$''$ size, oversampling the 2.7$''$ width of the spatial point response function measured on calibration stars, which allows for artifact-free image transformations.

For the northern portions of the mapped region, surface features were readily visible in the SOFIA/FORCAST data, making their association with known features straightforward. In the near-polar, southern regions (in particular, south of -85$^\circ$ latitude), the sky-plane view is highly foreshortened. To map the data onto selenographic coordinates, we used the images taken with the guide camera throughout the period of observation. The sharpest of the guide camera images taken during each FORCAST exposure was compared to a Lunar Reconnaissance Orbiter (LRO) image with matching illumination for the observation time. Together with the known location of the FORCAST slit in the guide camera pixels, we determined the lunar coordinates of each pixel along the FORCAST grism slit, for each exposure in each tile. This mapping enables identification of features even close to the south pole.

\begin{figure}
\includegraphics[width=8in]{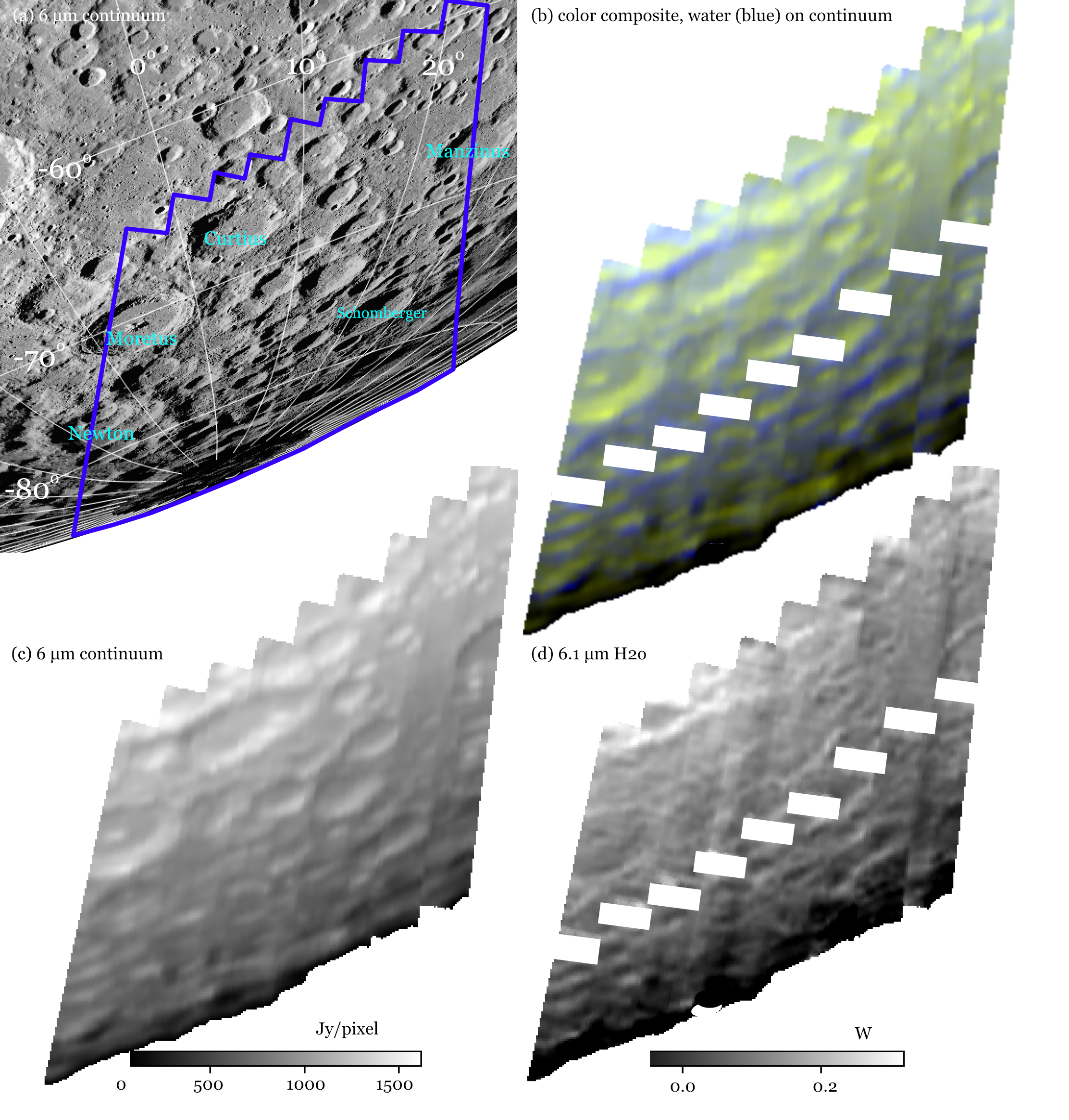}
\caption{{\it (a)} 
{\bfc Lunar Reconnaissance Orbiter Wide Angle Camera image of the southern limb of the Moon, oriented
with celestial pole at the top as appropriate for 2022 Feb 17, with lunar coordinate grid overlaid and an
outline of the region observed with SOFIA (blue lines).}
{\it (b)} Color rendition of SOFIA water and continuum emission. The color image combines the 6 $\mu$m continuum surface brightness (green) and the 6.1 $\mu$m water feature strength $W$ (blue). The  diagonal empty regions on the water image mask a partially transparent defect on the surface of the detector that leads to significantly higher noise. 
{\bfc 
{\it (c)} The 6 $\mu$m continuum surface brightness, together with color bar in Jy/pixel, for pixel size 0.768$"$. 
{\it (d)} The 6.1 $\mu$m integrated water band strength, $W$, defined in the text.
}
}
\label{fig:mooncolor}
\end{figure}

\section{Distribution of Water Emission in the Southern Hemisphere}

Figure~\ref{fig:mooncolor} shows the sky-plane H$_2$O map for the observed region, which spans from Moretus to Manzinus craters on the west and east, and from Curtius to the lunar limb (including the South Pole) from north to south. The H$_2$O feature is {\bfc strong} only in the western and northern portions of the field. The eastern half of the region has only weak H$_2$O, with an apparent negative feature on the easternmost side. Recalling the reference scheme described above, this means that the eastern side of the observed region has less H$_2$O than the Mare Fecunditatis reference, at the time of observation.

The H$_2$O emission is highest in the northwestern portion of the observed field, with distinct spatial structures. The spatial structure has some correspondence to lunar topography, but there is no one-to-one relationship. Most notable is a “wet ridge” that runs across the northern edge of Curtius crater and extends well beyond it to the western edge of the map. Also notable is water around the rim of Moretus crater. The water emission is just outside the bright crater seen in the continuum. 

\begin{figure}
\plotone{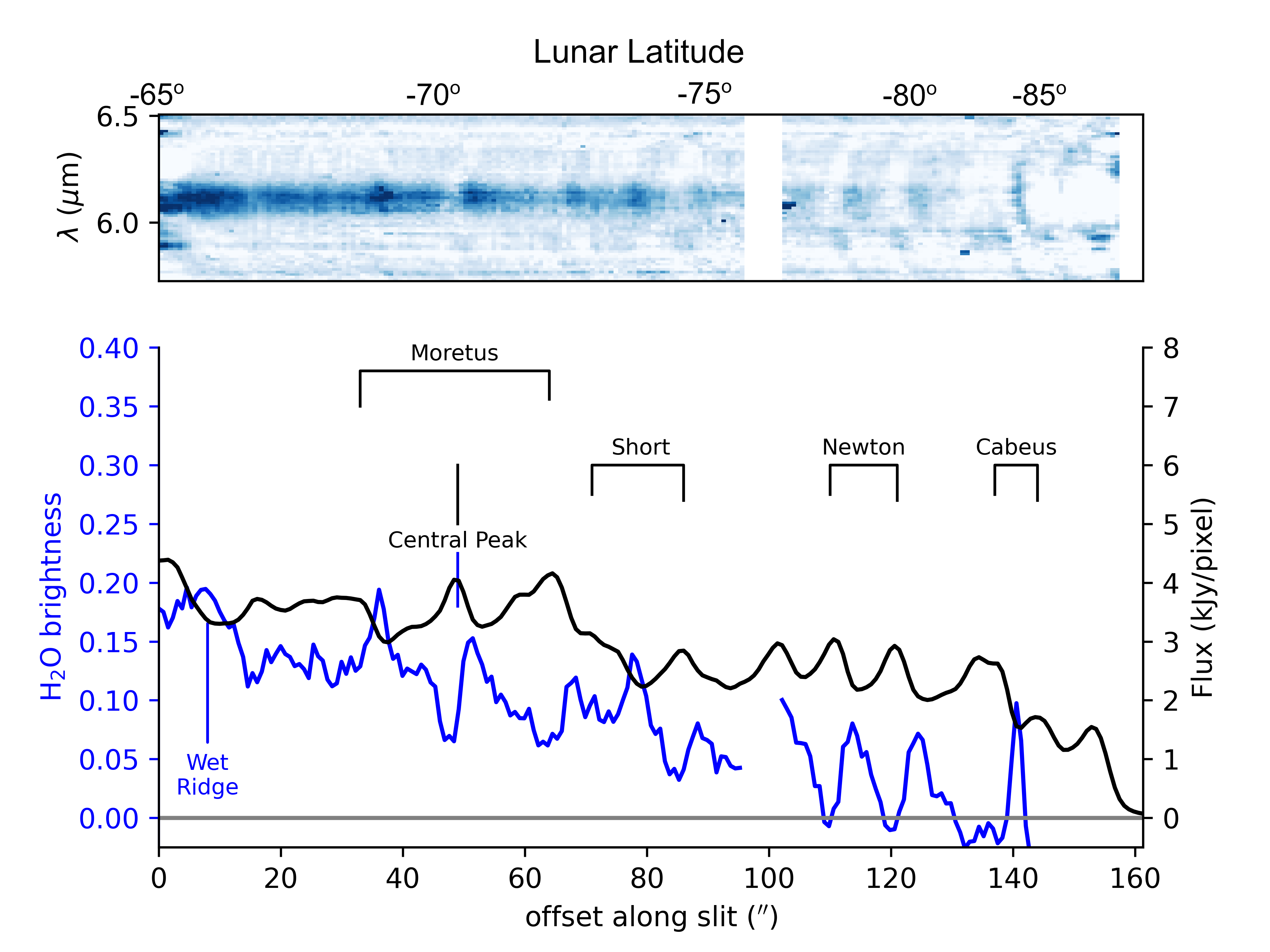}
\caption{Water profile through Moretus. The {\it top panel}, aligned with the lower panel along the horizontal axis, shows the two-dimensional spectrum from a single pointing of the spectrometer. {\bfc The shades of blue indicate the emissivity, relative to the continuum fit, from 0.997 (no color) to 1.019 (dark blue). } The thick, horizontal dark band is the 6.1 $\mu$m water feature. 
The {\it lower panel} shows spatial profiles of the 6.1 $\mu$m feature summed brightness W (blue) and the continuum (black). The profiles extend from the northernmost edge of Fig.~\ref{fig:mooncolor}, just right of the left-hand edge, traversing several labeled features including the Wet Ridge of high H$_2$O content, and the central peak of Moretus crater. 
The spatial structures are significant local enhancements in H$_2$O content, and they are superposed on a gradual trend that decreases from north to zero (immeasurably low) at the limb beyond the permanently shadowed crater Cabeus. In the top panel, horizontal stripes are electronic artifacts on the detector triggered by saturated pixels, and a small white patch near 101$"$ offset and 6.08 $\mu$m wavelength is due to a {\bfc defect} on the detector surface (mentioned in the Fig.~\ref{fig:mooncolor} caption).}\label{fig:spslice}
\end{figure}

Figure~\ref{fig:spslice} shows the H$_2$O {\bfc 6.1} $\mu$m feature strength along a slice running from the top to bottom of the map, beginning at lunar longitude,latitude of -3$^\circ$,-64$^\circ$ and ending at the lunar limb near -130$^\circ$,-89$^\circ$. 
The residual spectra were summed between 6.01 and 6.19 $\mu$m to generate integrated H$_2$O band strengths for each spectrum. The dimensionless feature brightness was calculated from the observed spectrum, $F_\nu$, by summing the excess emissivity over the channels between 6.01 and 6.19 $\mu$m: 
\begin{equation}
W = \Sigma (F_\nu/B_\nu - 1),
\end{equation}
where $B_\nu$ is the baseline continuum fit. The spectral channel width was 0.012 $\mu$m. Thus a 1\% amplitude emissivity excess spanning 0.12 $\mu$m would have $W$=0.1. An estimate of the water concentration (relative to the reference) can be made using the empirical calibration derived from observation of glass beads and radiative transfer modeling \citep{honniball_molecular_2021}, which yields water concentrations ({\bfc in units of} parts per million, or $\mu$g/g) approximately $900 \times W$. {\bfc In this paper we use the purely empirical measure $W$, and we address 
trends rather than the absolute quantity of water. Using $W$ can underestimate the water abundance if the spectral feature shifts in wavelength or is wide enough to spill out of the 6.01--6.19 $\mu$m sum. 
Inspecting the variation of feature center and width, the summed wavelengths should include essentially all of the emission.
Also note that $W$ is a lower limit because the water feature
from the reference position was removed in the calibration process. }

The 6.1 $\mu$m feature is {\bfc strong} in the northern part of the mapped region, with a gradual trend that decreases approaching the South Pole. High-amplitude fluctuations are superposed on this trend variation, beginning with the `wet ridge', then 
the northern rim of Moretus, then a dip at the northern face of the Moretus central peak and a peak at the southern side of the Moretus central peak, then the southern rim of Moretus, both rims of Short crater, then a decrease to near-zero values with peaks at rims of Newton and possibly Cabeus craters. The northern edges of crater-related features in the H$_2$O image are all shifted southward from the actual crater rims in reflected light or in 6 $\mu$m continuum emission. 

\begin{figure}
\plotone{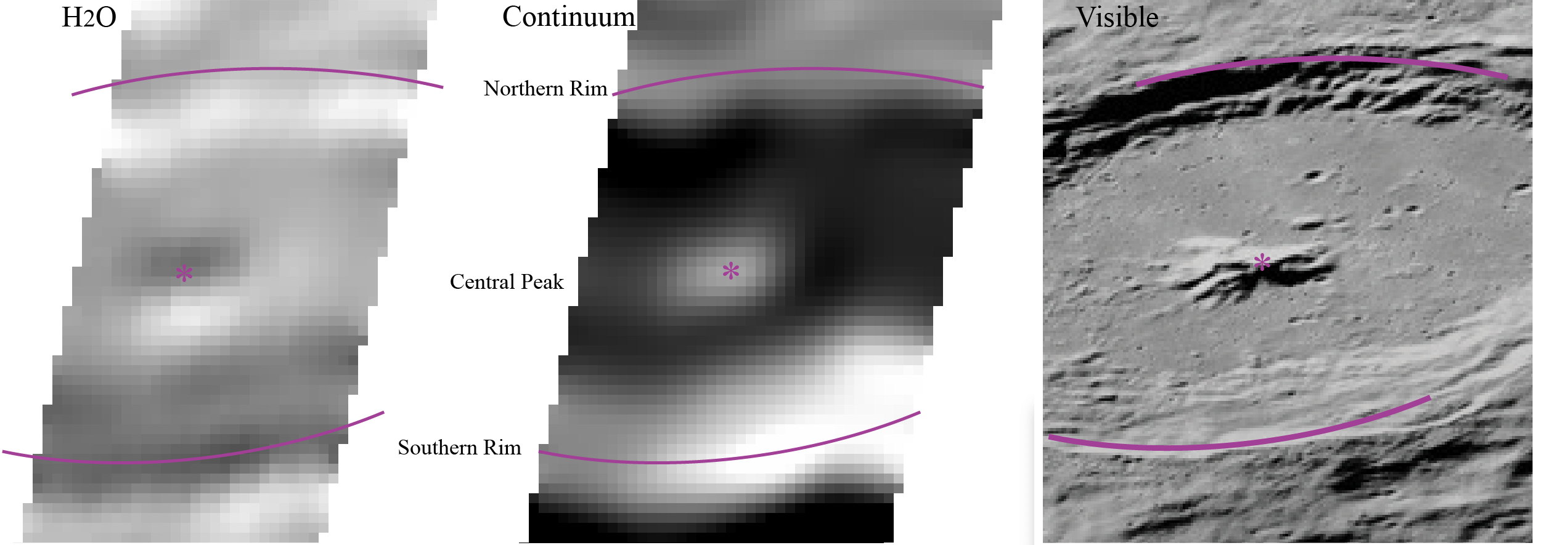}
\caption{Detail of water emission  at  Moretus. These SOFIA images of 6.1 $\mu$m H$_2$O emission (left) and 6 $\mu$m continuum (center) are centered on Moretus crater. A visible-light image (right) is from Chang’e 2, {\bfc using the Virtual Lunar Atlas \citep{chevalley_pchevvirtualmoon_2023}, at approximately the same illumination conditions (by coincidence). }
The purple asterisk indicates the top of the central peak in the crater, and purple arcs are guides to the locations of the north and south rims. These same markers are copied over to the left-hand panel. Both the rims and the central peak are shifted southward in the H$_2$O image compared to the continuum image, consistent with higher H$_2$O abundance on the sides of topographic features that are less illuminated. }\label{fig:moretus}
\end{figure}

The central peak of Moretus also has a distinct water pattern, with emission on the southern face and a deficit of emission on the northern slope (Fig.~\ref{fig:moretus}). The SOFIA observation was made near Full Moon, so the center of solar illumination is {\bfc north} of the region. The H$_2$O on the more-illuminated side is significantly decreased, while that of the shaded, southern face has a significant peak of H$_2$O emission. This pattern is reminiscent of the well-known effect that the north face of a mountain in the northern hemisphere has more snow than the southern face.
{\bfc The  surface temperature, measured by Diviner and compiled by \citet{williams_global_2017} at 15 km sampling, shows that the southern face of the Moretus central peak is anomalously
cold. In particular, the duration of its heated day is much shorter than the surroundings. At 3 hours after noon in lunar time, the Diviner sample closest to the south face of Moretus central peak is already down to 177 K, in contrast to the sample 15 km away, where the temperature is still at its noon-time value (298 K). 
The significantly higher water abundance measured by SOFIA, faster cooling measured by Diviner, and reduced insolation evident from LRO/LOLA altimetry 
all suggest the poleward, southern face of the central peak of Moretus is a distinct environment from its surroundings and worthy of further investigation.}

\section{Distribution of Water Emission near the South Pole}

\begin{figure}
\plotone{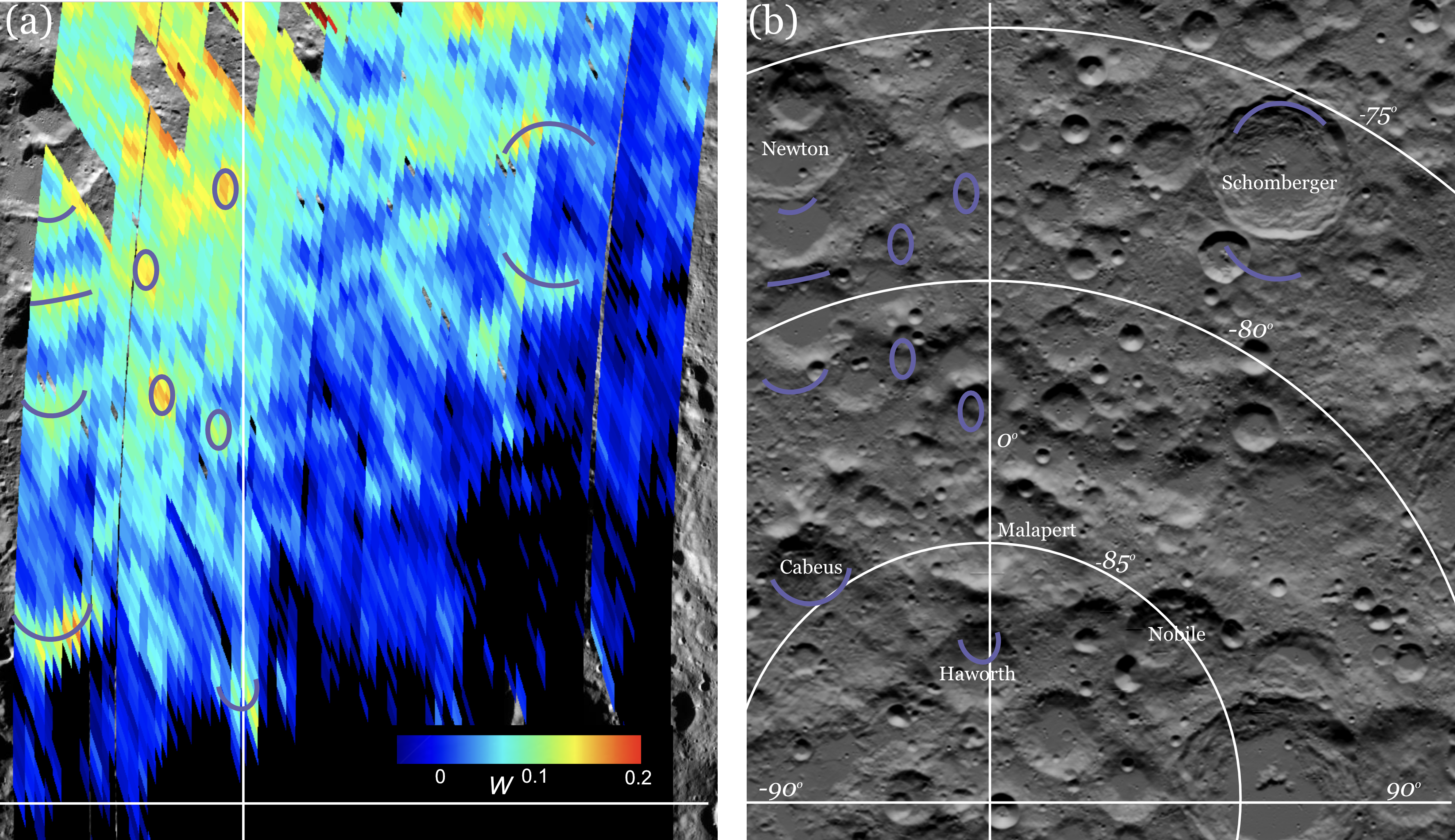}
\caption{Distribution of water near the lunar south pole. {\it (a)} In false color, a stereographic projection of the 6.1 $\mu$m H$_2$O emission. 
{\it (b)} In greyscale, an LRO image of the exact same area. 
{\bfc The longitude 0$^\circ$ and 90$^\circ$ meridians are shown in both panels,
and latitude circles are labeled panel {\it (b)}.}
The pixels in the SOFIA image are elongated due to reprojection from the foreshortened, Earth-centered view of the pole, which was close to the lunar limb but just-visible at the time of the SOFIA observation. 
{\bfc Arcs (grey in panel {\it a} and purple in panel {\it b}) }
delineate some of the water emission features. Particularly noteworthy in this projection is bright water emission from just inside the northern rims of Cabeus and Haworth. The water emission appears to be the crater floors just at the inner base of the northern rims; this also occurs at Shomberger.  }
\label{fig:polar}
\end{figure}

On the date of the SOFIA observation, the lunar south pole and a small portion of the far side were visible from Earth. To show the distribution of water closer to the south pole, Figure~\ref{fig:polar} shows an orthographic projection. Significant local concentrations of water emission follow the northern rim of Cabeus, the center of Haworth, and the southern rim of Nobile. Compared to the trends seen at higher latitudes, the Cabeus northern rim is consistent with what was seen at Curtius, Moretus, and other craters. {\bfc Cabeus and Haworth} craters contain permanently shadowed regions \citep{cisneros_permanetly_2018} and include evidence of water ice, notably Cabeus from the LCROSS impact \citep{colaprete_detection_2010}, so the near-polar craters may have larger regions that are amenable to procuring and/or retaining molecular water \citep{farrell_spillage_2015}. The molecular water emission closely follows the location of shadows in the LRO image, even though the truly shadowed regions are too cold to contribute significantly
to the mid-infrared FORCAST images. The strong peak of molecular water on the southern rim of Nobile may be of interest to inform the VIPER rover \citep{colaprete_volatiles_2020}, which is set to explore that region in late 2024. 
{\bfc The details of
the distribution of 6.1 $\mu$m water emission and its relation to the shadowed regions requires a detailed study,
taking into account the effects of local topography and highly oblique viewing that controls which surfaces are actually seen in the SOFIA observations.}

\section{Discussion of Potential Relation to the Source of Lunar Water}
The lunar south polar region, from Curtius to Manzinus and down to the South Pole shows a significant 6.1 $\mu$m emissivity feature that is uniquely associated with H$_2$O based on its central wavelength and width. The presence of the feature confirms the prior detections in this region of the Moon \citep{honniball_molecular_2021}, and the large-scale contiguous mapping now shows systematic spatial variations associated with lunar surface features. 

An overall trend of decreasing H$_2$O toward the south pole is seen in the data. 
{\bfc This trend can be
compared to prior observations, taken at somewhat different locations and lunar times of day. \citet {honniball_molecular_2021} showed the water feature was stronger ($W\sim .4$) at latitude -75$^\circ$ and 
decreased to a plateau ($W\sim 0.2$) at latitude -55$^\circ$; the range of values is similar to what is seen in the map reported here, but the trend versus latitude is different.
\citet{honniball_regional_2022} showed in a larger regional map that the water feature was stronger ($W\sim 0.3$) for the higher-flux surface
at latitudes around -60$^\circ$ and decreased (to $W\sim 0.1$) toward the lower-flux surface at latitudes around -80$^\circ$.
The latitude trends from the limited prior observations are different from the the trend derived from the new data reported here, but the prior observations did not go to the pole and covered different longitudes. The different latitude trends could be due to local differences due to different parts of the lunar surface that was sampled,
or due to the different lunar time. 
The subsolar longitude was $-12^\circ$ for the observations presented in this paper, 
compared to $+21^\circ$ for the prior regional map \citep{honniball_regional_2022}; at Moretus crater, that
means the lunar time of day was just 0.4 hr after noon for the new observations and 1.8 hr before noon for the prior 
regional map. It is possible that the lunar time of day plays a significant role in the mobility of water at high latitudes.
Further observations with SOFIA were performed of both the southern and northern polar regions, which will allow this issue to be studied in more detail in the future.}
A decreasing trend with latitude could potentially be related to the lower flux of solar wind and meteoroids that contribute to lunar hydration \citep{klima_remotely_2017}. The solar wind flux has a cosine-of-latitude variation, with relatively little reaching the pole. The meteoroid influx on the Moon is of order 2 tons/day mostly from Jupiter-family comets \citep{carrillo-sanchez_cosmic_2020}, which could coat the entire surface of the Moon with a depth of 7 cm if spread uniformly with a volume averaged mass density of 1 g~cm$^{-3}$ over 4 Gyr. The influx of sporadic meteoroids has a {\bfc H}elion source of small particles ($\beta$-meteoroids) moving radially outward from the Sun, but most of the mass flux is in the apex/anti-apex sources, due to larger particles on elliptical orbits arising from comets and asteroids \citep{wiegert_dynamical_2009,pokorny_dynamical_2014}. 
A tenuous lunar dust atmosphere driven by meteoroid bombardment has been associated with the apex source \citep{szalay_annual_2015}. Both solar wind implantation and sporadic meteoroid bombardment provide a source function for water onto the Moon that should decrease toward the poles. That trend is consistent with the decrease in 6.1 $\mu$m H$_2$O we see in the SOFIA observations presented in this paper, which must be noted as covering only a fraction of the lunar surface. The trend is opposite from the equator-to-pole OH+H$_2$O trend seen with M3 \citep{li_direct_2018}, but it only covers a limited latitude range. Meteoroid streams associated with individual comets on inclined orbits \citep{brown_meteoroid_2008} give rise to significant increases when the Moon traverses them, which would contribute cometary water input at higher latitudes. 
Indeed, the existence of water produced by impacts from meteoroids in known streams was discovered by LADEE, though it can be interpreted as pre-existing lunar water liberated by the impacts rather than exogenic delivery \citep{benna_lunar_2019}. Analysis of the LCROSS plume indicates a cometary origin for the volatiles in the permanently shadowed region of Cabeus \citep{mandt_exogenic_2022}, where we also detect water with SOFIA. Further SOFIA observations covering a wider range of latitudes are being analyzed and will address latitude variations as well as localized emission in the 6 $\mu$m feature.

\section{Conclusions}

The presence of significant small-scale variations in water content associated with topographic features (Fig.~\ref{fig:mooncolor}), with amplitude comparable to an underlying smooth latitude variation (Fig.~\ref{fig:spslice}), shows that local processes on the lunar surface {\bfc (at least partially to topography)} control the distribution of water as much as any exogenic input. 
The systematic shift of water distribution away from solar illumination shows that the water has higher abundance just south of the topographic maxima. As the mapped region is in the far southern hemisphere, the shift is consistent with more water in the more-shadowed regions hidden for longer portions of the day by local topographic maxima. Fig.~\ref{fig:moretus} shows that the exposed, inner side of the southern rim of Moretus has a weak water signal. The shadowed, outer part of the southern crater rim (which rises 2.1 km above the mean surface and is much higher than the northern rim, which rises only 1.1 km) has a relatively high water abundance. Similarly, there is a water signal associated with the northern rim, from the shadowed, inner side of that rim.

A distinct H$_2$O emission pattern {\bfc appears} near the Moretus central peak (Fig.~\ref{fig:moretus}). Both the northern and southern faces of the Moretus central peak have extensive steep (60$^\circ$) slopes \citep{barker_new_2016}, extending vertically to 2.5 km above the crater floor. The base of the central peak has a size comparable to the physical resolution of the data (4.5 km), and the elongated shape is consistent with foreshortening of the crater at -70$^\circ$ latitude. This means that some of the spatial variation is smoothed out by the observing technique. Nonetheless the data reveal a clear excess of H$_2$O on the southern face as well as a clear deficit (relative to surroundings) in H$_2$O on the northern face of the central peak. The excess emission is likely related to insolation, which is only for a much shorter portion of the lunar day on the southern side. The decrease of water on the northern side of the central peak, below the water level at the floor of the crater, indicates that the steep northern slopes lose their water at a significantly higher rate, due to their long exposure to sunlight each lunar day. The steep slopes of the central peak could also lead to different regolith properties and depth, but the strong distinction between water on the north and south sides suggests {\bfc that} insolation is a more significant effect in controlling the water distribution.

The presence of localized enhancements in water abundance may support the hypothesis that at least part of the water 
produced by solar wind and meteoroid bombardment on the lunar surface may subsequently migrate toward the poles \citep{schorghofer_theoretical_2017,hendrix_diurnally_2019}, {\bfc  that localized topography and roughness may
play a role in water retention on the lunar surface \citep{davidsson_implications_2021}, and that colder portions
of the polar regions may retain water even if not permanently shadowed \citep{kloos_temporal_2019,hayne_micro_2021}.}
Both the observations from Deep Impact \citep{sunshine_temporal_2009,laferriere_variability_2022} and  SOFIA  
show higher H$_2$O abundance at where the surface is locally colder, and where migrating water may collect in cold traps and be reduced at high temperature on a diurnal cycle.
{\bfc The observational results presented in this paper pertain to the distribution of water traced by the 6.1 $\mu$m emissivity feature, 
which  arise from water molecules very near the surface but could trace a more pervasive source.
In addition to the localized enhancements of H$_2$O in colder regions that are detailed 
in this paper, there remains widespread H$_2$O see as a 6.1 $\mu$m emissivity peak
\citep{honniball_regional_2022} and OH+H$_2$O seen as 3 $\mu$m reflectivity dip \citep{wohler_time--daydependent_2017} across the lunar surface, which may
arise from extensive amounts of water in glasses or minerals.}

\begin{acknowledgments}
Based in part on observations made with the NASA/DLR Stratospheric Observatory for Infrared Astronomy (SOFIA). SOFIA was jointly operated by the Universities Space Research Association, Inc. (USRA), under NASA contract NNA17BF53C, and the Deutsches SOFIA Institut (DSI) under DLR contract 50 OK 0901 to the University of Stuttgart. 
Financial support for the observational and theoretical work in this paper was provided by NASA through award \#09\_0171 issued by USRA.
\end{acknowledgments}

\facility{SOFIA} 
\software{Virtual Moon Atlas \citep{chevalley_pchevvirtualmoon_2023},
Lunar QuickMap \citep{malaret_lunar_2022},
astropy \citep{astropy_collaboration_astropy_2022},
matplotlib \citep{hunter_matplotlib_2007}
}

\bibliography{references}

\end{document}